\documentclass[preprint,amsmath,amssymb,aps,prb]{revtex4-1}

\usepackage{graphicx}
\usepackage{dcolumn}
\usepackage{bm}
\usepackage{array}

\usepackage[T1]{fontenc}
\usepackage{subcaption}
\begin{document}

\title{Inverse pressure-induced Mott transition in TiPO${}_4$\\}

\author{H. Johan M. J\"onsson\textsuperscript{1}}
\author{Marcus Ekholm\textsuperscript{1}}
\author{Maxim Bykov\textsuperscript{3,5}}
\author{Leonid Dubrovinsky\textsuperscript{5}}
\author{Sander van Smaalen\textsuperscript{3}}
\author{Igor A. Abrikosov\textsuperscript{1,4}}

\affiliation{{}\textsuperscript{1}Theoretical Physics Division, Department of
Physics, Chemistry and Biology (IFM)}
\affiliation{{}\textsuperscript{2}Swedish e-Science Research Centre (SeRC),
Link\"oping University, SE-581 83, Link\"oping, Sweden}
\affiliation{{}\textsuperscript{3}Laboratory of Crystallography, University of
Bayreuth, 95440 Bayreuth, Germany}
\affiliation{{}\textsuperscript{4}Materials Modeling and Development Laboratory,
 National University of Science and Technology `MISIS', Moscow, 119049, Russia}
\affiliation{{}\textsuperscript{5}Bayerisches Geoinstitut, University of Bayreuth,
95440 Bayreuth, Germany}

\begin{abstract}
TiPO$_4$ shows
interesting  structural and magnetic properties as temperature and pressure are
varied, such as a spin-Peierls phase transition and the development of
incommensurate modulations of the lattice.
Recently, high pressure experiments for TiPO$_4$ reported two new structural
phases appearing at high pressures, the so-called phases IV and V [M. Bykov et al.,
Angew. Chem. Int. Ed. 55, 15053]. The latter was shown to
include the first example of 5-fold O-coordinated P-atoms in an inorganic
phosphate compound.
In this work we characterize the electronic structure and other
physical properties
of these new phases by means of \emph{ab-initio} calculations, and investigate the structural transition.
We find that the appearance of phases IV and V coincides with a collapse of
the Mott insulating gap and quenching of magnetism in phase III as
pressure is applied. Remarkably, our calculations show that in the high
pressure phase V, these features reappear, leading to an
antiferromagnetic Mott insulating phase, with robust local moments.

\end{abstract}

\maketitle

\section{\label{sec:Intro}Introduction}\noindent
Modern advances in experimental high-pressure techniques have lead to numerous
discoveries of new phenomena and previously unknown phases of materials
\cite{jrn:dubrovinsky15,jrn:bykov18,jrn:troyani2016}.
Recently, experiments based on synchrotron x-ray diffraction in diamond anvil
cells \cite{jrn:bykov16} found
two new high-pressure phases of TiPO$_4$. One of these new phases, phase V, showed the
first experimental confirmation of 5-fold O-coordinated P-atoms in an inorganic
phosphate compound.

\begin{figure*}[ht]
    \centering
    \begin{subfigure}{0.49\textwidth}
        \includegraphics[trim={3cm 4cm 40cm 11cm},width=\linewidth,clip]{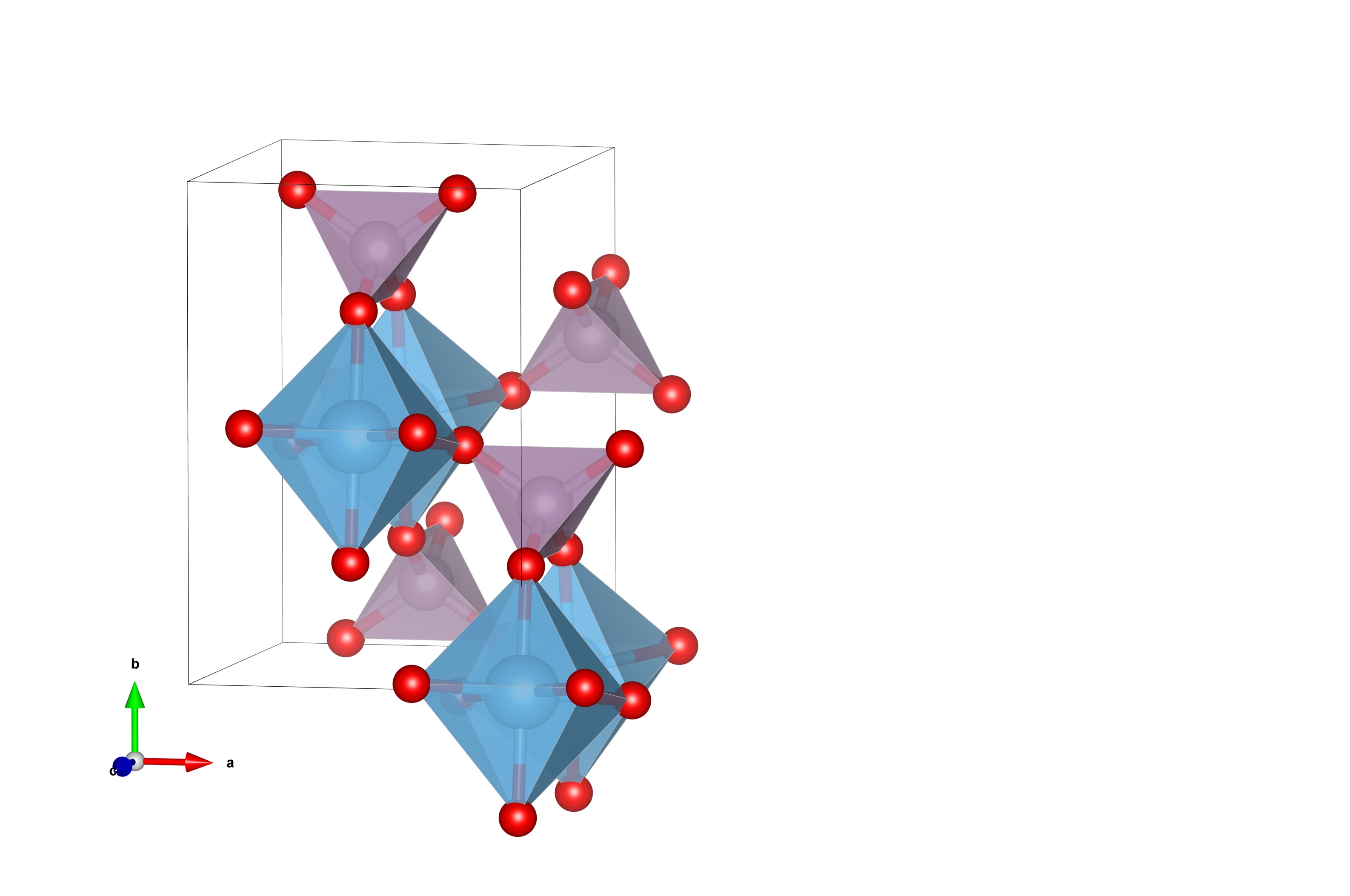}
        \caption{}
        \label{fig:cryststruct-I}
    \end{subfigure}
    \begin{subfigure}{0.49\textwidth}
        \includegraphics[trim={7cm 5cm 20cm 0cm},width=\linewidth,clip]{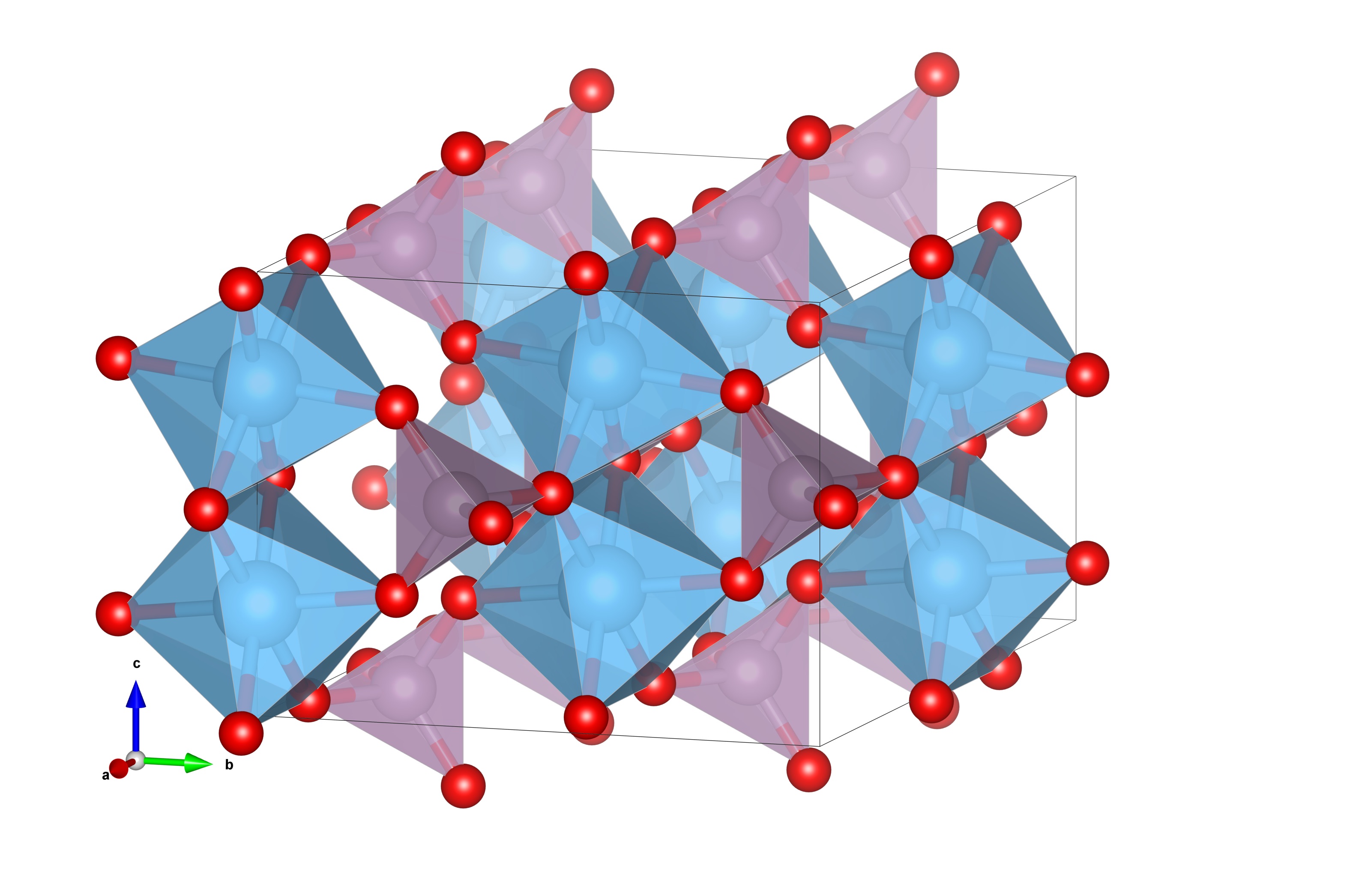}
        \caption{}
        \label{fig:cryststruct-III}
    \end{subfigure}
    \begin{subfigure}{0.49\textwidth}
        \includegraphics[trim={0cm 0cm 0cm 0cm},width=\linewidth,clip]{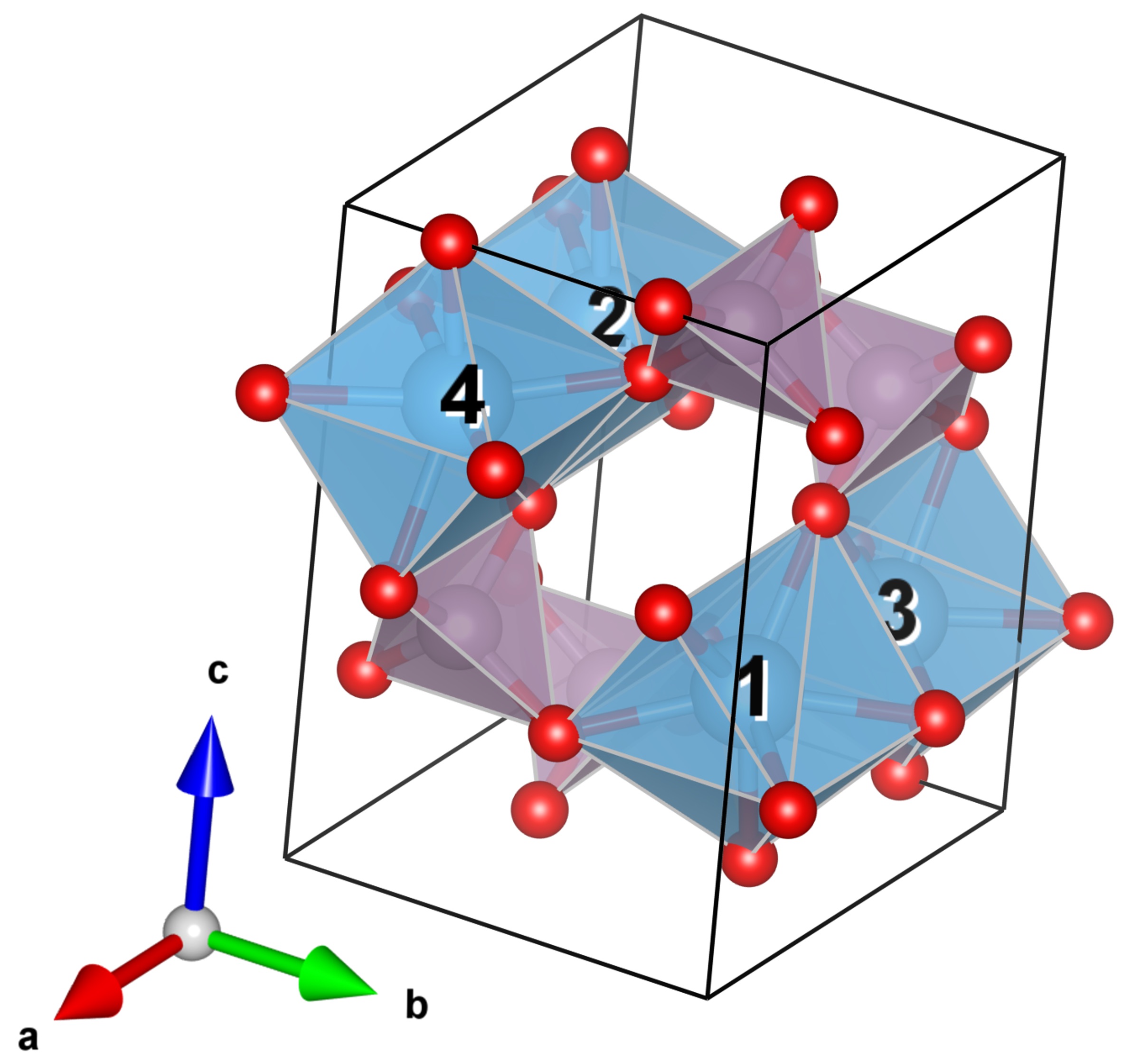}
        \caption{}
        \label{fig:cryststruct-IV}
    \end{subfigure}
    \begin{subfigure}{0.49\textwidth}
        \includegraphics[trim={0cm 0cm 0cm 0cm},width=\linewidth,clip]{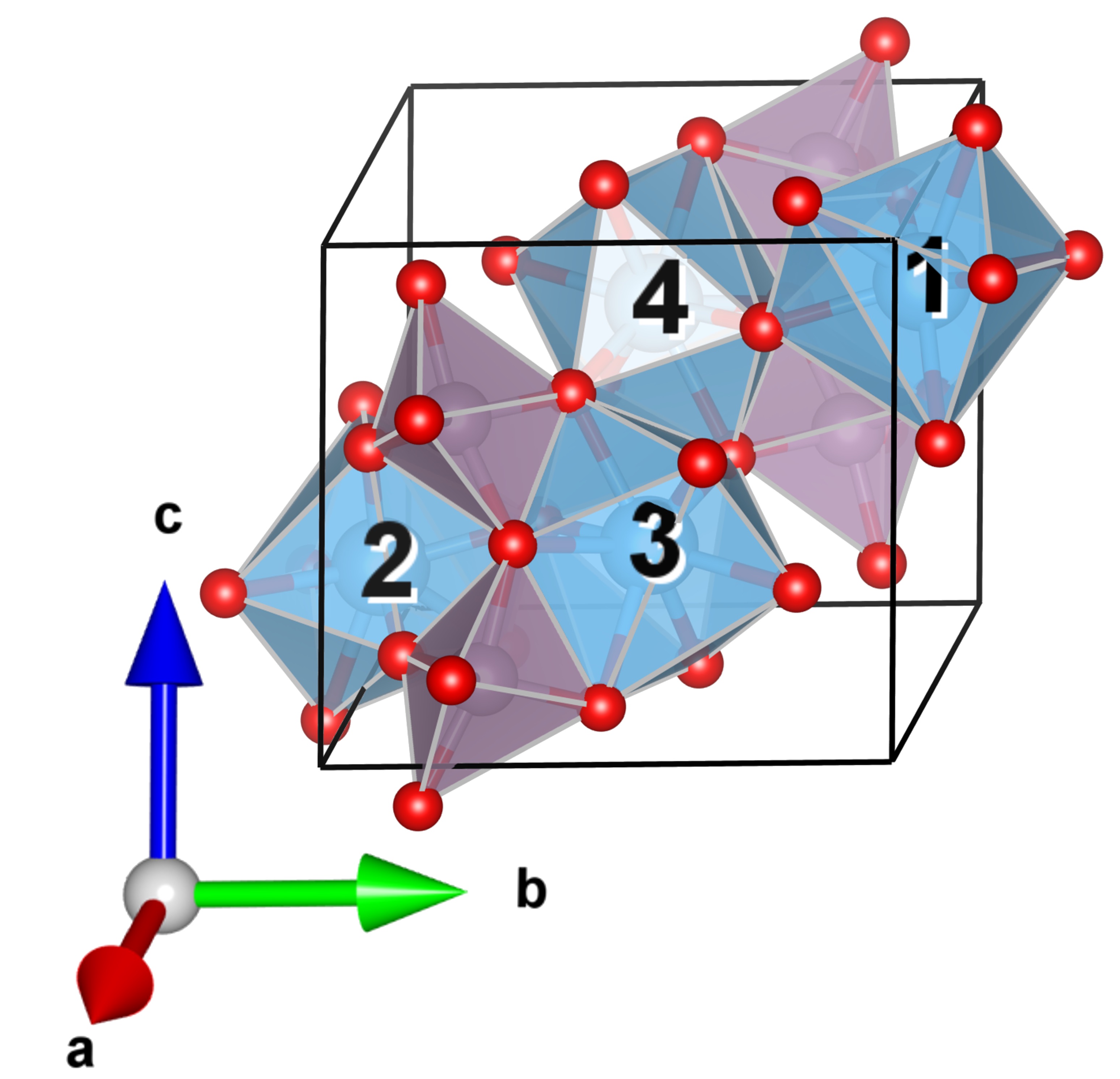}
        \caption{}
        \label{fig:cryststruct-V}
    \end{subfigure}
    \caption{Crystal structures observed in TiPO$_4$ under increasing
    compression a) phase I,  b) phase III, c) phase IV and d) phase V. Ti atoms are shown in blue, P are purple and O are red. Also shown in blue
    are the Ti-O polyhedra and the P-O polyhedra are purple.}
    \label{fig:cryststructs}
\end{figure*}

At ambient conditions, TiPO$_4$ crystallizes in the orthorhombic
CrVO$_4$-structure (space group
$Cmcm$) which features  TiO$_6$ and PO$_4$ polyhedra (Fig.
\ref{fig:cryststruct-I}), and is referred to as
phase I\cite{jrn:bykov16}.
The edge-sharing TiO$_6$ octahedra form quasi-one-dimensional (1D) magnetic
chains along the c-axis.
By means of nuclear magnetic resonance measurements in combination with
\emph{ab-initio} calculations, magnetic coupling along these chains has been
determined to be antiferromagnetic (AFM)\cite{jrn:law11}.
TiPO$_4$ has attracted attention as being one of only a few inorganic materials
undergoing a spin-Peierls (SP) transition with a particularly high critical
temperature of $T_{SP} = 74$ K \cite{jrn:law11,jrn:wulferding13,jrn:bykov13},
which can be compared to  $T_{SP} = 14$ K in
CuGeO$_3$, and 34 K in NaV$_2$O$_5$\cite{jrn:hase1993,jrn:isobe1996}.
The temperature of the SP transition in TiPO$_4$ appears to be sensitive
 to the applied pressure. Ambient temperature compression to 4.5 GPa leads to
 an incommensurately modulated structure (phase II), that is seen upon
 cooling below 111 K at ambient pressure. The commensurate spin-Peierls phase
 III (Fig. \ref{fig:cryststruct-III}) becomes stabilized at 7.3
 GPa\cite{jrn:bykov16} at room temperature.

Phase III can be regarded as a fourfold superstructure of phase I, and features the same
 TiO$_6$ and PO$_4$ polyhedra.
However, above 46 GPa, the diffraction pattern has been reported to change drastically, revealing a
 structural transition into two coexisting phases, called IV
  and V (Figs. \ref{fig:cryststruct-IV} and \ref{fig:cryststruct-V})\cite{jrn:bykov16}.
Phase IV is similar to phase III but with a denser hexagonal packing of the
O-atoms. This means that the TiO$_6$ octahedra become highly distorted, leading to
TiO$_7$ capped trigonal prisms while the PO$_4$ tetrahedra are retained.
Phase V has monoclinic P$2_1 / m$ symmetry, with interlinked TiO$_7$ and the
unprecedented PO$_5$ polyhedra.

Little is known about these high-pressure phases, due to the difficulties
associated with conducting in-situ measurements in diamond-anvil cells. In this
work, we investigate the electronic structure of the novel high pressure phases
of TiPO$_4$. In particular, we focus on the relative
stability of phases IV and V, as well as orbital and magnetic ordering, and
demonstrate the exciting Mott physics that occurs in highly compressed TiPO$_4$.
Our experiments reveal that upon compression the band gap in phase III
shrinks, as indicated by a reddening of the color of the samples. At the
transition to phases IV and V the sample contains a mix of phases IV and V, and
its color turns black. \emph{Ab-initio} calculations show that the gap is indeed closed in
phases III and IV at high pressure. However, phase V features a sizeable Mott gap and magnetic
moments.

\begin{figure}
    \begin{subfigure}{0.49\linewidth}
        \includegraphics[width=\linewidth]{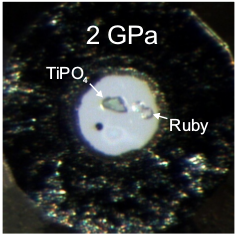}
        \caption{}
        \label{fig:exp-phases-2}
    \end{subfigure}
    \begin{subfigure}{0.49\linewidth}
        \includegraphics[width=\linewidth]{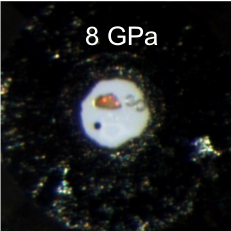}
        \caption{}
        \label{fig:exp-phases-8}
    \end{subfigure}
    \begin{subfigure}{0.49\linewidth}
        \includegraphics[width=\linewidth]{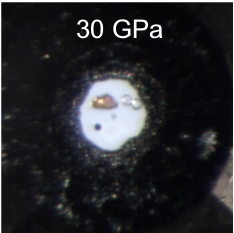}
        \caption{}
        \label{fig:exp-phases-30}
    \end{subfigure}
    \begin{subfigure}{0.49\linewidth}
        \includegraphics[width=\linewidth]{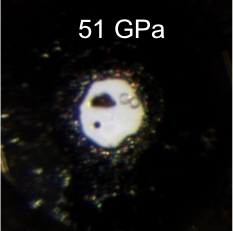}
        \caption{}
        \label{fig:exp-phases-51}
    \end{subfigure}
    \caption{Sample of TiPO$_{4}$ at different pressures. Note the colour change
    as pressure is increased.}
    \label{fig:exp-phases}
\end{figure}

\section{\label{sec:Experiment}Experimental details}\noindent
A single crystal of TiPO$_4$ from the same synthetic batch that
was used for the high-pressure studies in Ref. \citenum{jrn:bykov16} was placed
inside a sample chamber in a BX90 diamond anvil cell. Neon
was used as a pressure-transmitting medium, while a small
chip of ruby was used as a pressure standard\cite{jrn:mao86}. The sample was
compressed in a few steps up to 51 GPa to track the color
change of the crystal upon compression.

\section{\label{sec:Method}Computational details}\noindent
Calculations were performed using the projector augmented wave
\cite{jrn:blochl94,jrn:kresse99}
 (PAW) method implementation in the Vienna Ab-initio Simulations Package
\cite{jrn:vasp1,jrn:vasp2} (VASP). For
the antiferromagnetic calculations we employed a plane wave energy cut-off of
520 eV.
The calculations employed the local density approximation\cite{jrn:perdew81}
(LDA) with an added on-site effective
interaction parameter, U (LDA + U) as parametrized by Dudarev et al.\cite{jrn:lda+u}
Calculations of phase I employed a k-point mesh of 10x8x10 k-points using the Monkhorst-Pack
scheme\cite{jrn:MonkhorstPack},
with a total of 200 k-points in the irreducible Brillouin zone. For phase III we
employed a k-point mesh of 5x7x11 k-points, resulting in a total of 112 k-points
in the irreducible Brillouin zone. For the structural optimization of phases IV and V
we employed k-point meshes of 12x10x8 and 7x4x5, resulting in 240 and 70
k-points in the respective irreducible Brillouin zones. For the calculations of
the electronic structure of phase V a refined k-point
mesh of 9x7x5, giving 158 k-points in the irreducible Brillouin zone, was used.

Calculated energies and volumes were fitted to the third order Birch-Murnaghan
equation of state (EOS)\cite{jrn:birch47}.
We chose the value U$=2.0$ eV because it adequately reproduces the experimental
equation of state, as demonstrated below, as well as the quite complex pressure
dependence of the crystal structure parameters of phase III, as will be
presented elsewhere. 

\section{\label{sec:Res}Results}\noindent
Fig. \ref{fig:exp-phases} shows the change in color of the TiPO$_4$ crystal on
compression in a diamond anvil cell. Upon phase
transformation to phase III, the color of the crystal
changes from green (Fig. \ref{fig:exp-phases-2}) to orange (Figs.
\ref{fig:exp-phases-8}, \ref{fig:exp-phases-30}), while the
transition to phases IV and V results in the sample turning black.

\begin{figure}
    \includegraphics[width=\linewidth]{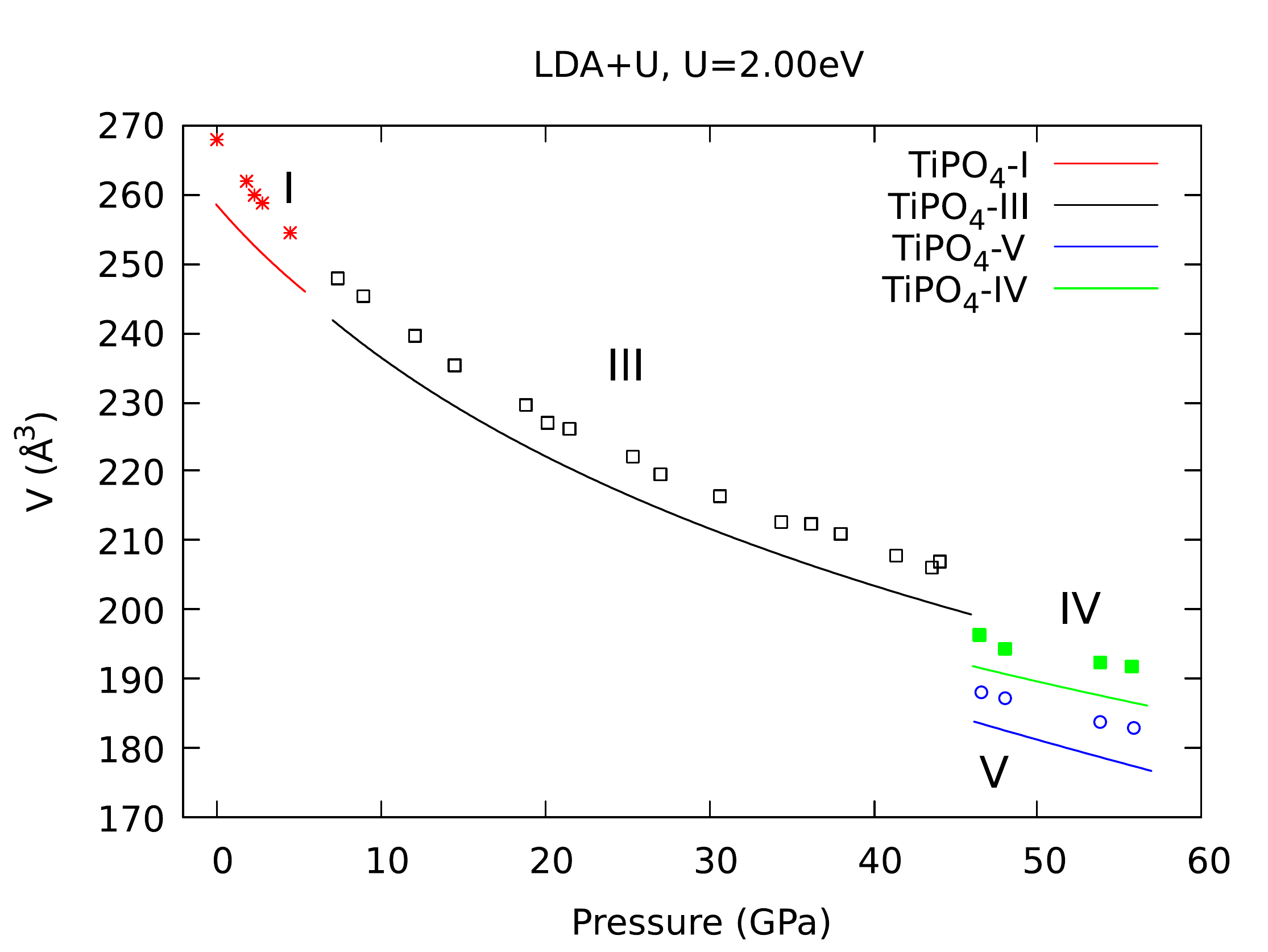}
    \caption{Equation of state for TiPO${}_4$. Points are experimental
    data\cite{jrn:bykov16}. Solid lines show EOS calculated in this work.}
    \label{fig:eos}
\end{figure}

In Fig. \ref{fig:eos} we show the obtained volume-pressure relation for
phases I, III, IV and V. Phase I was set up in the same AFM configuration used in
earlier calculations\cite{jrn:lopezmoreno12}.
We considered phase III in the AFM configuration suggested by experiments \cite{jrn:bykov16}.
For phases IV and V we compare the calculated enthalpy, $H = E(V) + PV$, for
four different magnetic configurations;
ferromagnetic $\uparrow\uparrow\uparrow\uparrow$, as well as three antiferromagnetic
configurations, $\downarrow\uparrow\uparrow\downarrow$,
$\downarrow\uparrow\downarrow\uparrow$ and $\uparrow\uparrow\downarrow\downarrow$
(the order of Ti atoms is shown in Fig. \ref{fig:cryststructs}). For phase IV,
all configurations lie within 2 meV/atom from each other, with the AFM configuration
$\downarrow\uparrow\downarrow\uparrow$ having the lowest enthalpy. In phase V, the difference
in enthalpy is at most 4 meV/atom, with the AFM
configuration $\downarrow\uparrow\uparrow\downarrow$ being the lowest. The
difference in enthalpy is very small compared to typical thermal energy at room
temperature ($kT \sim 20$ meV). This means that at room temperature, phases IV and V are most
likely in a paramagnetic state with disordered local moments.

There is a pressure-independent underestimation of the theoretical
unit cell volume of about 4\% compared to experiments. However, the overall
 behaviour of
the EOS is very well reproduced. Our results for $B_0$ agree well with previous results obtained for phase
I using the PBE functional, without on site interaction U
 \cite{jrn:lopezmoreno12}, however we obtain a lower value of $V_0$. This should
 be seen in the light that our calculations do
 not include thermal effects and that the PBE functional fails to reproduce an insulating
 ground state for phase III. The transition from phase III to phases
IV and V includes a decrease in the unit cell volume. In Table
\ref{tab:eqparam} we summarize the equilibrium parameters obtained from the
EOS fits for the different phases. Note that for phase IV and V
there is a decrease in equilibrium volume and a significant increase in bulk modulus.
Comparing phases III and V, there is a twofold increase of $B_0$ for the
high pressure phase.

\begin{table}
    \caption{Equilibrium parameters from calculations and experiment. Experimental
    values for phases I, II and III were determined in Ref. \citenum{jrn:bykov16} from a single fit.}
    \begin{tabular}{| l | l | l | l |}
        \hline
    	& V${}_0$ (Å\textsuperscript{3})  & B${}_0$ (GPa) & B${}'_0$ \\
        \hline
        Phase I & 258.56 & 92 & 6.5 \\
        \hline
        Phase III & 258.79 & 87 & 5.4 \\
        \hline
        \hline
        Experiment\cite{jrn:bykov16} (phases I, II and III)	& 267.86	& 72	& 6.5\\
        \hline
        \hline
        Phase IV & 236.46 & 133.11 & 4.8 \\
        \hline
        Phase V & 225.95 & 179 &  2.4 \\
        \hline
    \end{tabular}
\label{tab:eqparam}
\end{table}



\begin{figure}
    \includegraphics[width=\linewidth]{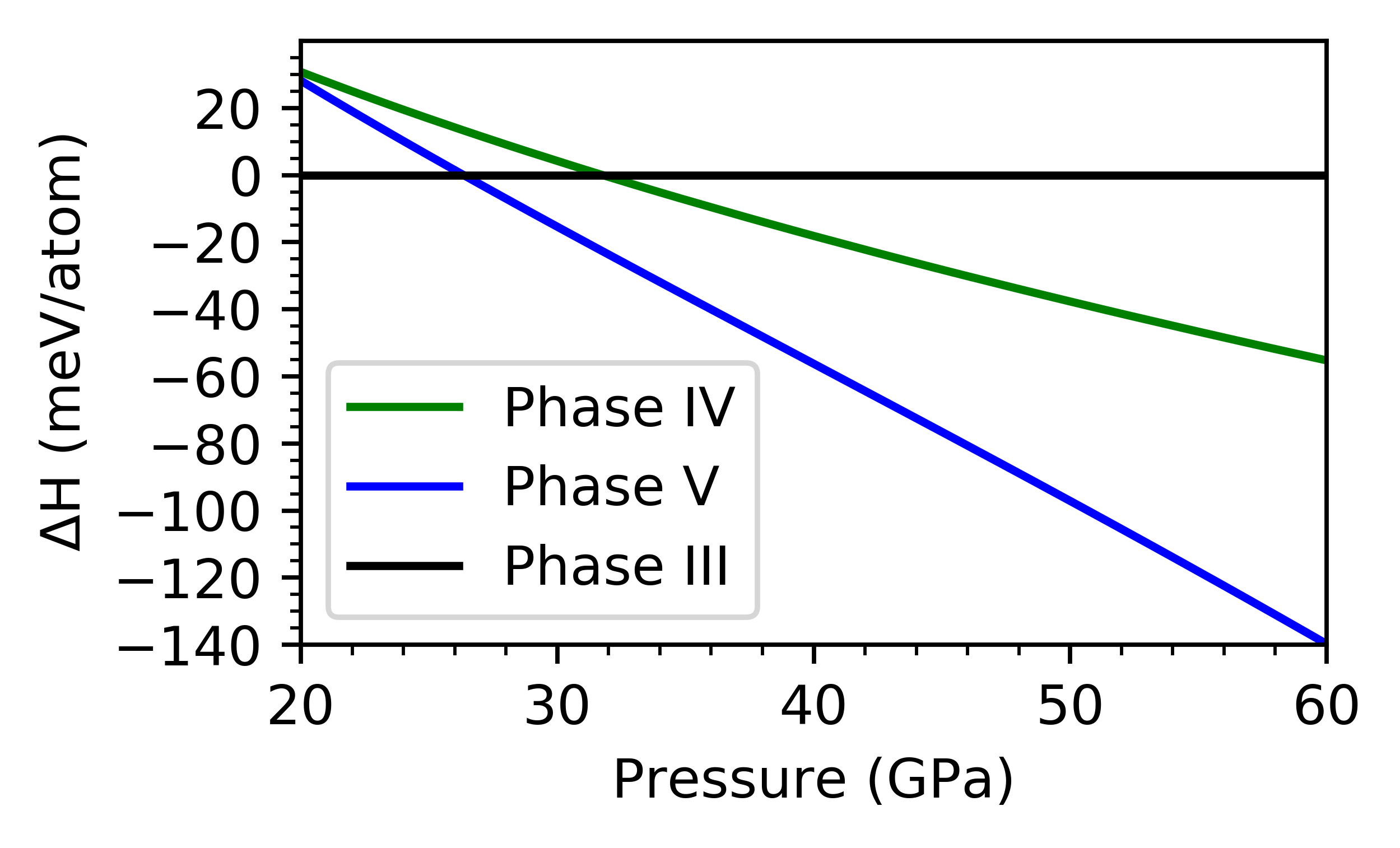}
    \caption{Enthalpy for phase IV, and V, relative to phase III
    as a function of pressure.}
    \label{fig:enth}
\end{figure}
In Fig. \ref{fig:enth} we show the enthalpies of phases IV and V relative to
phase III. We see that phase V becomes favorable over phase III
at around 25 GPa, and phase IV becomes favorable over phase III at around 33 GPa.
In (room temperature) experiments these two phases have
been reported to appear at around 45 GPa.
The underestimation of the transition pressure is expected, due to the systematic
underestimation of the volume in the EOS (the pressure is thus too
low at each volume).
As outlined in Ref. \citenum{jrn:bykov16} the transition from phase III
 to phase IV involves only small changes to the crystal
structure, whereas the transition from phase III to V involves substantial
rearrangement.
Interestingly, phase V is lower in enthalpy than phase IV in the entire pressure range.
Thus it appears that phase V is the thermodynamically stable phase at high
pressures, while phase IV is \emph{kinetically} stable, which explains the
coexistence of the two structures in the high-pressure experiment\cite{jrn:bykov16}.
Phase IV maintains the alternating Ti-Ti distances along the a-direction
observed in the SP phase III\cite{jrn:bykov16}.

\begin{figure*}
    \begin{subfigure}{0.45\linewidth}
        \includegraphics[width=\linewidth,trim={0 0 0 0}, clip]{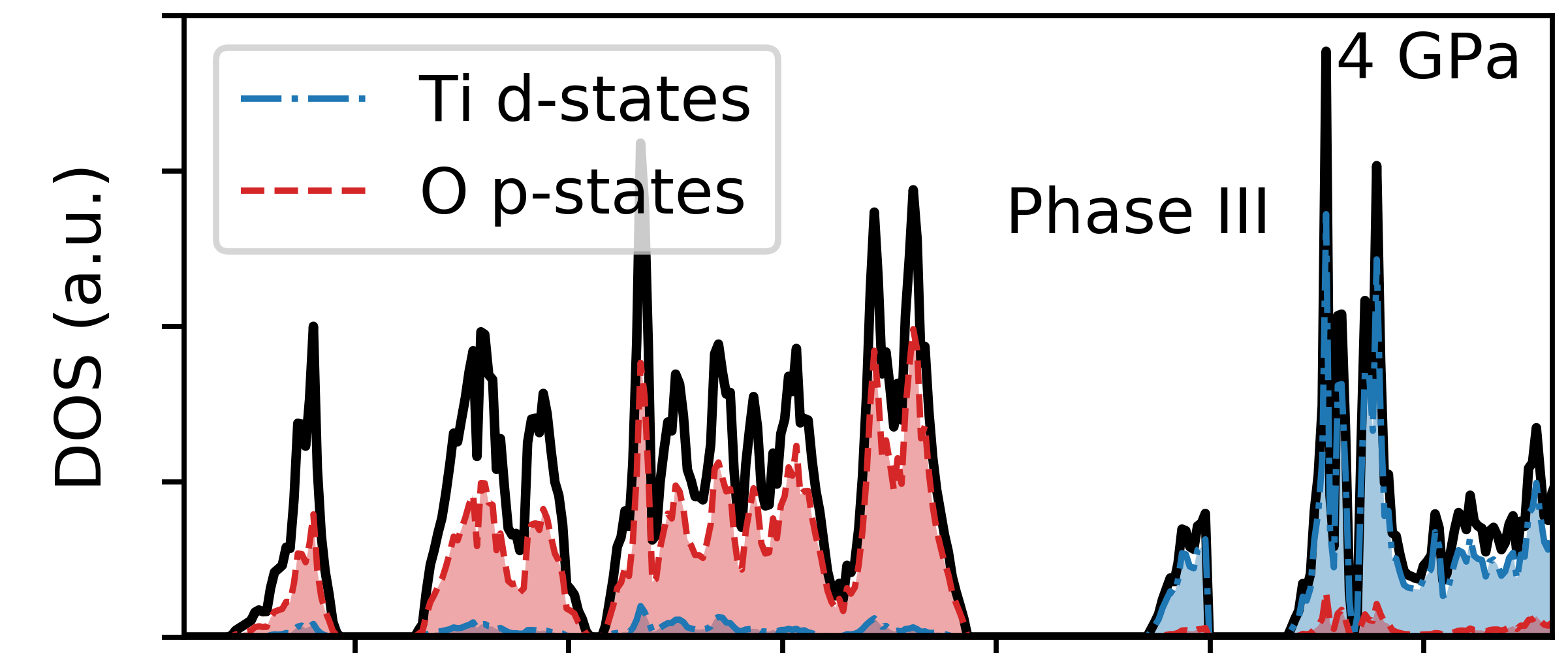}
        \caption{}
        \label{fig:IIILPDoS}
    \end{subfigure}
    \begin{subfigure}{0.45\linewidth}
        \includegraphics[width=\linewidth,trim={0 0 0 0}, clip]{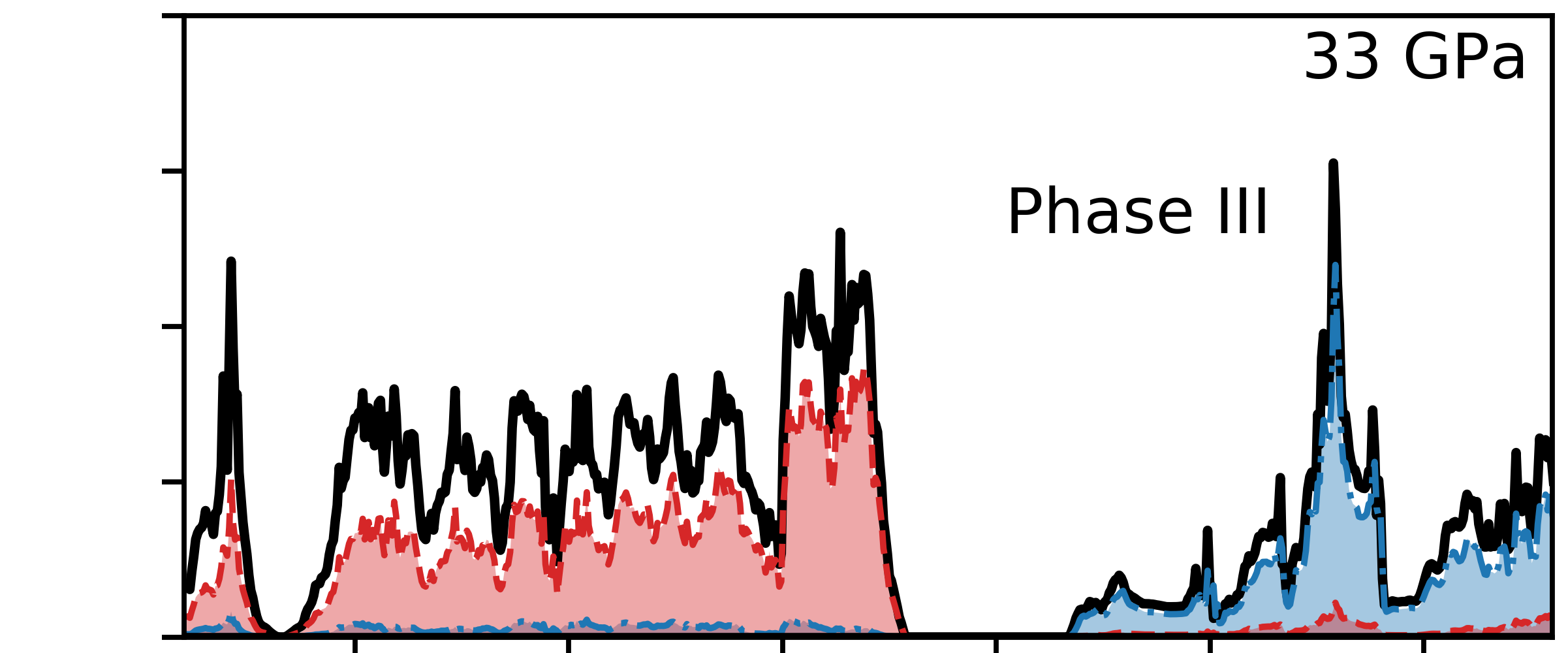}
        \caption{}
        \label{fig:IIIHPDoS}
    \end{subfigure}
    \begin{subfigure}{0.45\linewidth}
        \includegraphics[width=\linewidth,trim={0 0 0 0}, clip]{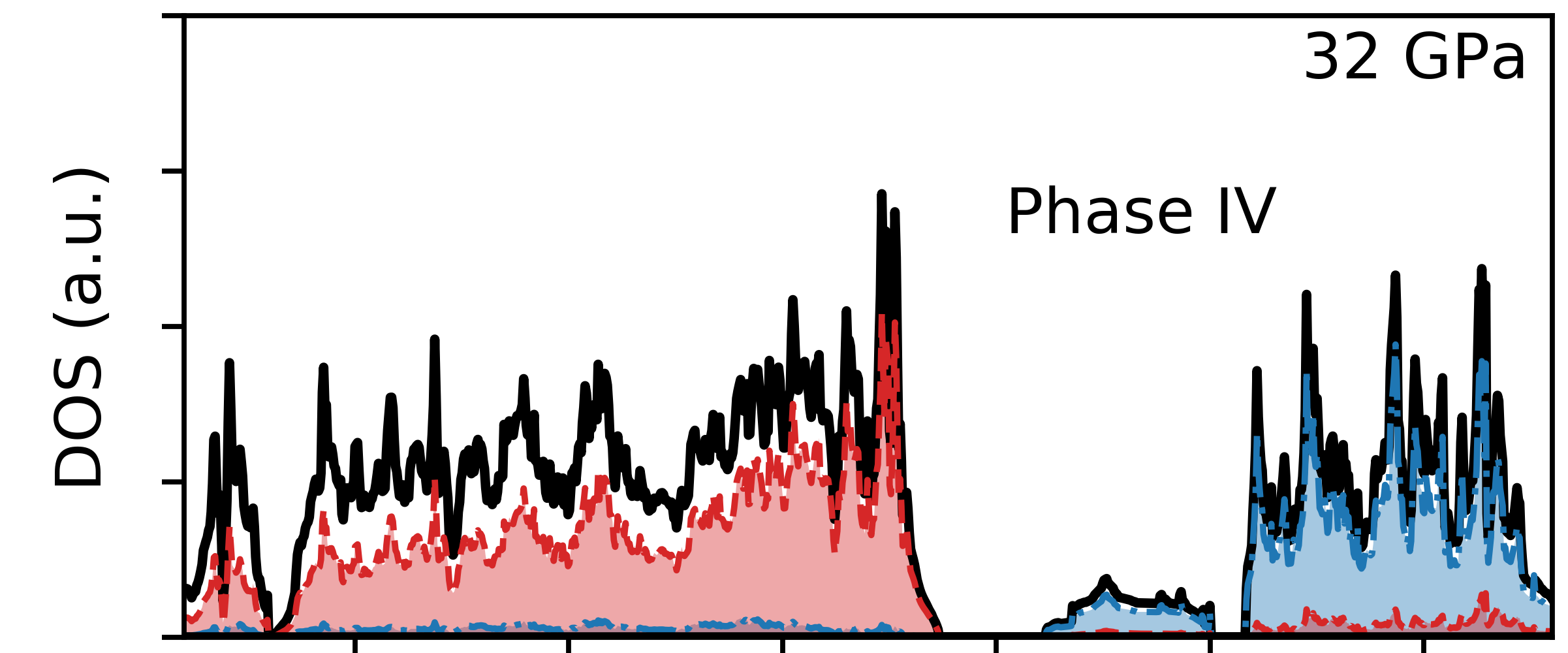}
        \caption{}
        \label{fig:IVLPDoS}
    \end{subfigure}
    \begin{subfigure}{0.45\linewidth}
        \includegraphics[width=\linewidth,trim={0 0 0 0}, clip]{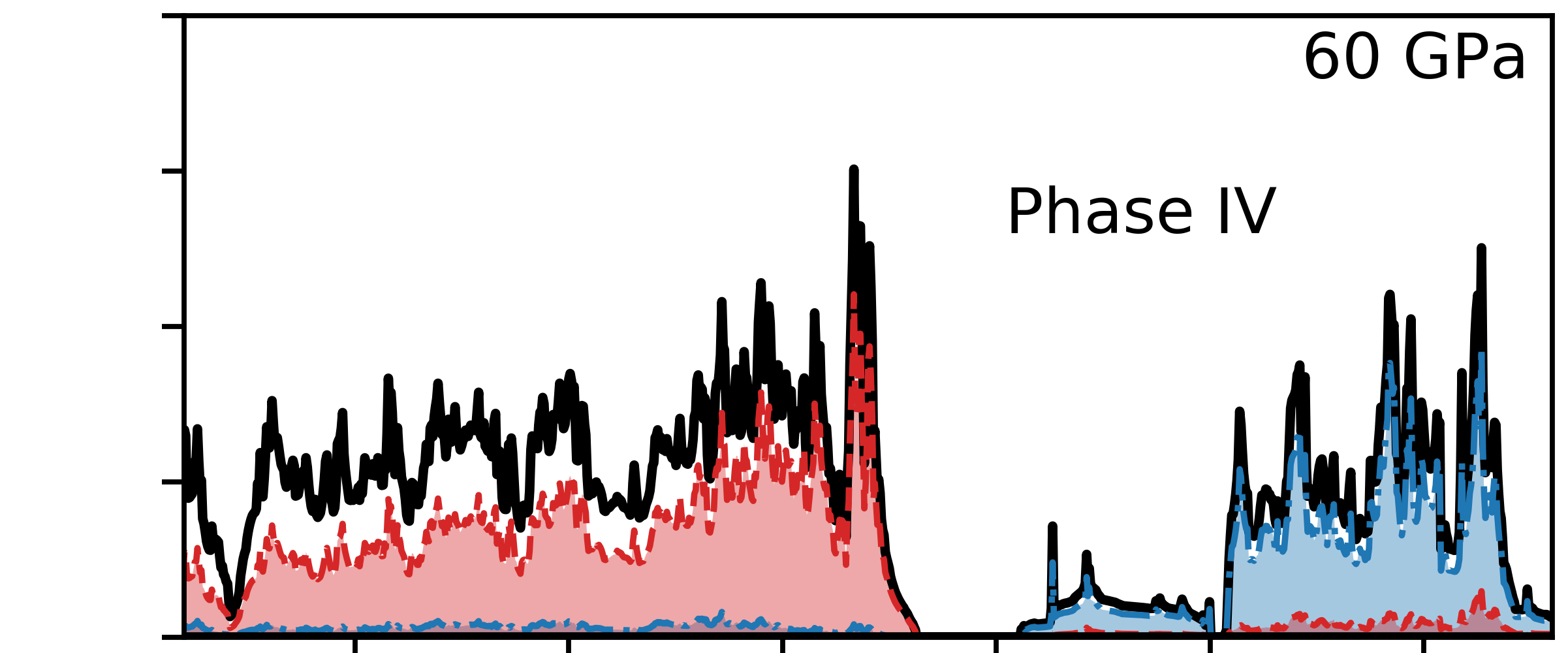}
        \caption{}
        \label{fig:IVHPDoS}
    \end{subfigure}
    \begin{subfigure}{0.45\linewidth}
        \includegraphics[width=\linewidth,trim={0 0 0 0}, clip]{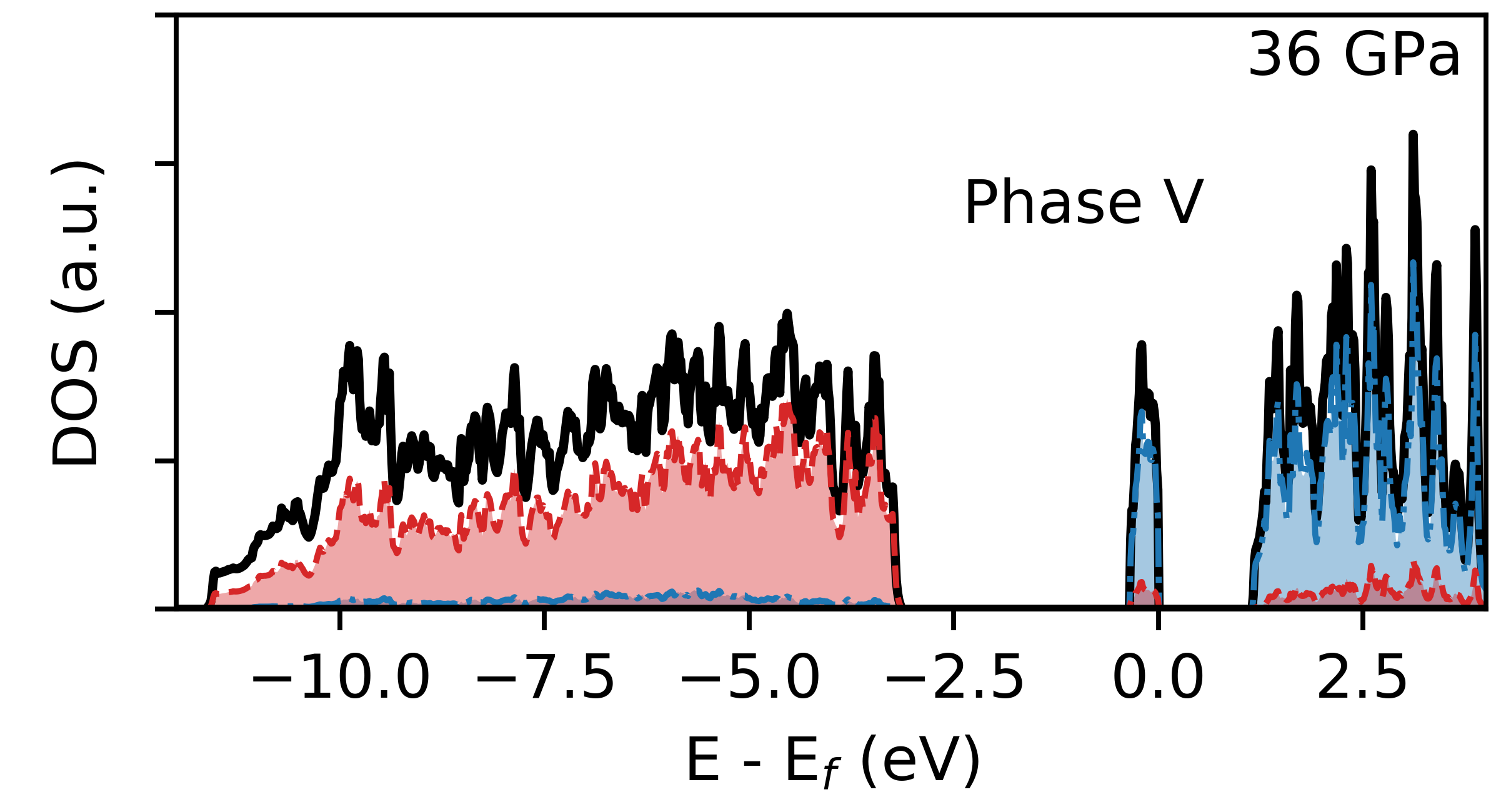}
        \caption{}
        \label{fig:VLPDoS}
    \end{subfigure}
    \begin{subfigure}{0.45\linewidth}
        \includegraphics[width=\linewidth,trim={0 0 0 0}, clip]{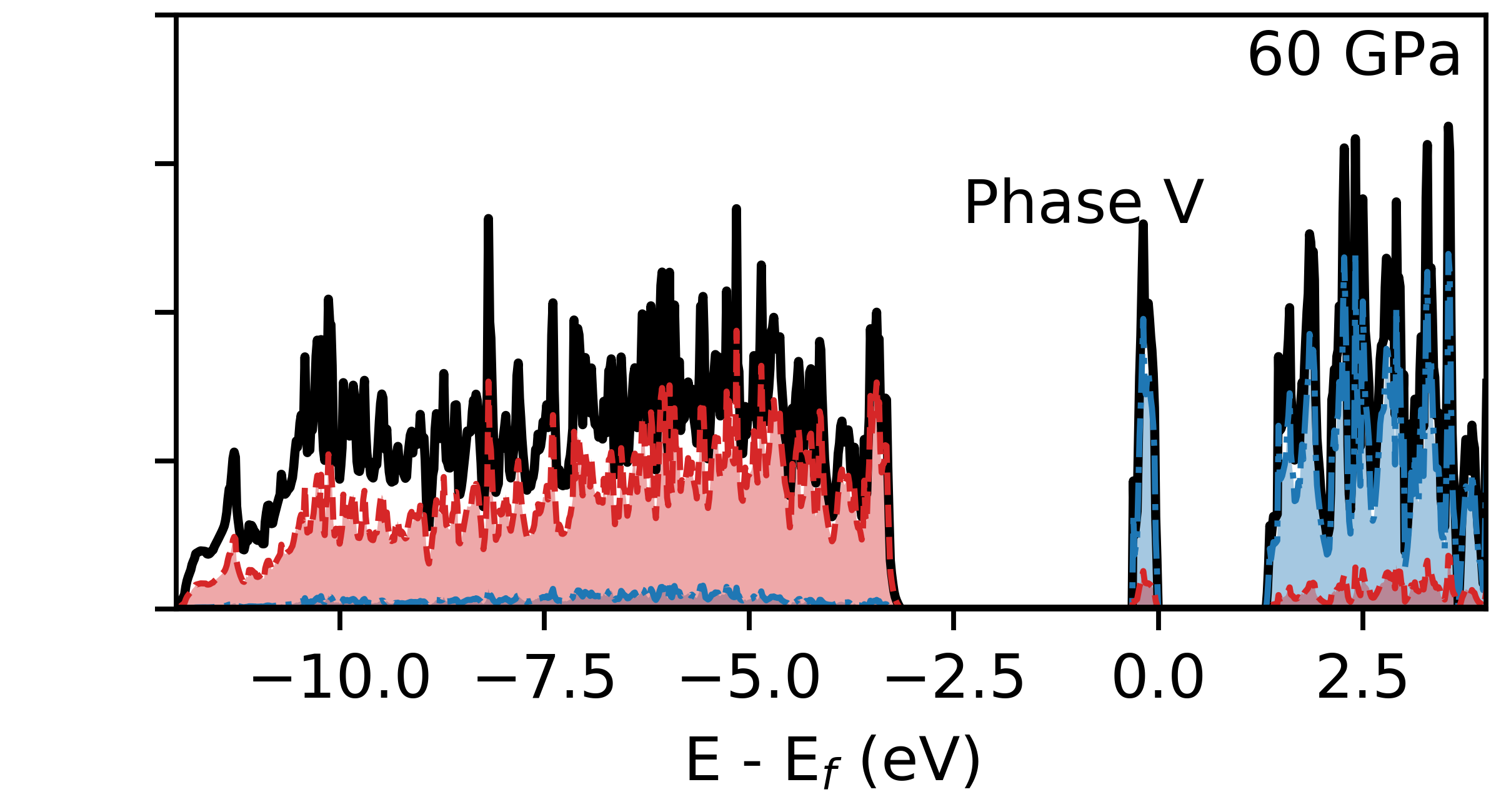}
        \caption{}
        \label{fig:VHPDoS}
    \end{subfigure}
    \caption{Black line refers to the total DOS of the unit cell. Blue and red
    refer to the site and orbital-projected DOS. Energy given relative to the Fermi energy (E$_F$).}
    \label{fig:DoSColl}
\end{figure*}

Phase III is stable for a large range of pressures. Figure
 \ref{fig:IIILPDoS} shows
its calculated density of states (DOS) at 4 GPa. A band gap between the occupied
 and unoccupied states is
clearly visible. This is in line with experimental observations
(Fig. \ref{fig:exp-phases}) that the samples are transparent at this pressure.
 As pressure is
increased the gap at the Fermi level shrinks, also in agreement with experiment.
 In
Fig. \ref{fig:IIIHPDoS} we see that at a pressure
around 33 GPa the gap in phase III is
completely closed. This pressure is lower than the experimental pressure at
which samples lose their transparency. However, it is the same pressure at which phases IV and V
become energetically favorable in calculations. Although calculations
underestimate the transition pressure we conclude that, in agreement with
experiment, the closing of the gap and the structural transition from phase III
to phases IV and V coincide.


\begin{figure}
    \includegraphics[width=\linewidth]{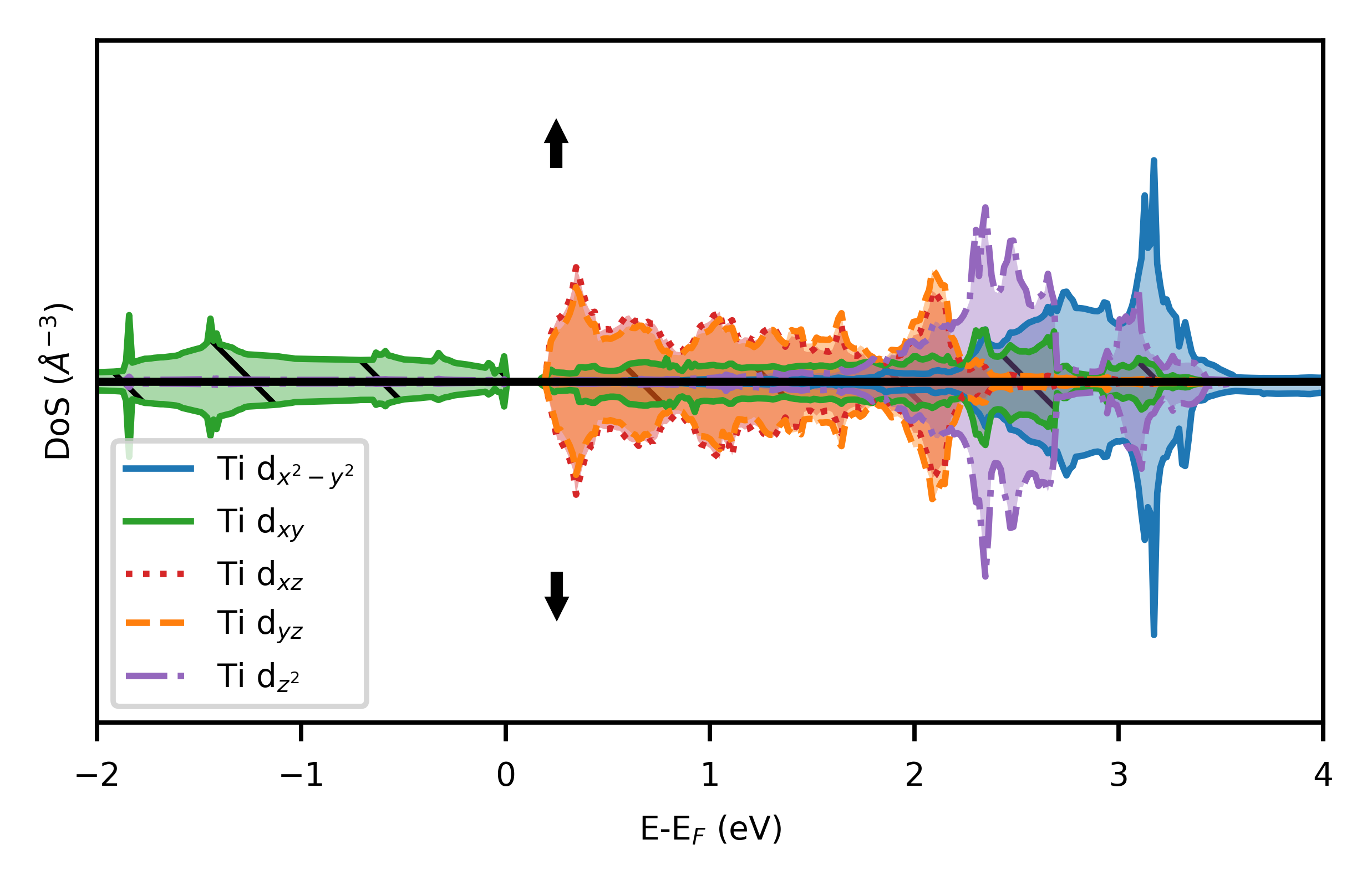}
    \caption{Calculated spin-resolved total DOS for one Ti atom in phase IV, at 60 GPa.}
    \label{fig:IVHPTiDoS}
\end{figure}

\begin{figure}
    \includegraphics[width=\linewidth,trim={0 0 0 0},clip]{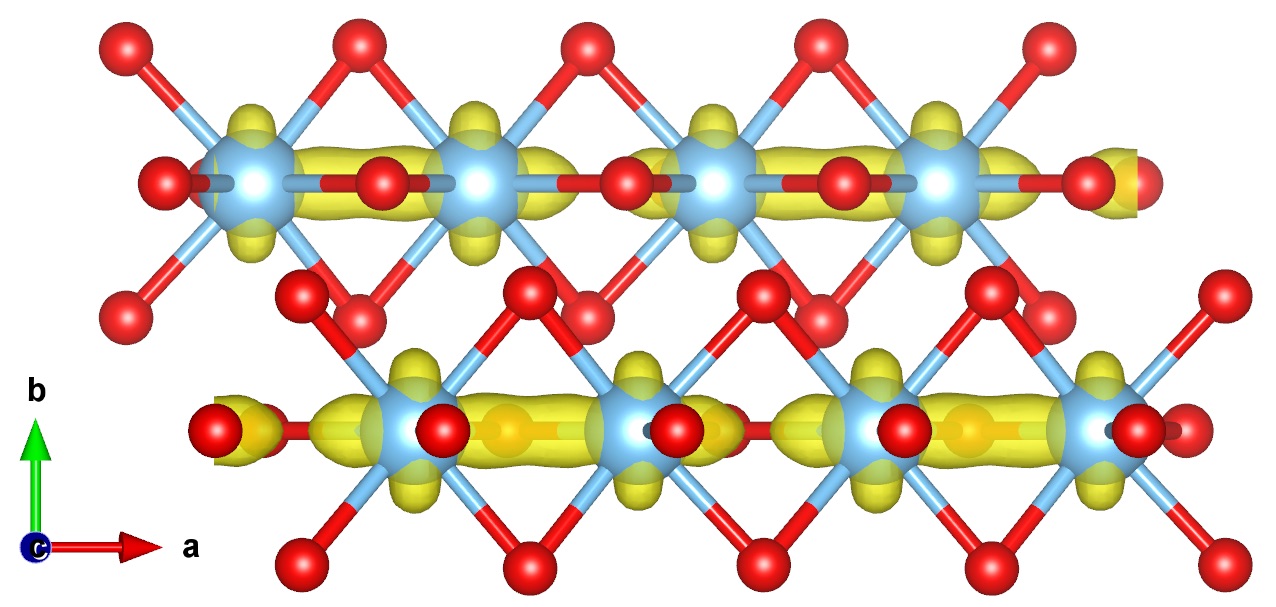}
    \caption{Partial charge density of Ti-d electrons in phase IV, at 60 GPa.
    The local $\bar{x}$, $\bar{y}$ and $\bar{z}$-axes have been chosen so that the $\bar{z}$ and $\bar{x}$-axes point
    towards oxygen atoms.}
    \label{fig:IVHPParChg}
\end{figure}

In Fig. \ref{fig:IVLPDoS} we show the calculated DOS for TiPO$_4$
phase IV at a pressure of 32 GPa. A very small band gap of $0.4$ eV is visible.
The low-binding energy part of the spectrum is of mainly Ti-d character,
corresponding to a singly occupied
d$_{xy}$ orbital, as shown in Fig. \ref{fig:IVHPTiDoS}.
This is
separated from a high-binding part of filled O-p states. In Fig.
\ref{fig:IVHPParChg} we show the occupied d-orbitals plotted in real space. There is
a clear overlap of bonding orbitals
in  Ti-chains along the a-direction, indicating direct exchange coupling.
 Note the separation of charge between
Ti-pairs, due to the alternating Ti-Ti distances.
In Fig. \ref{fig:IVHPDoS} we show the calculated DOS
of phase IV at a pressure of 60 GPa. The band gap has been reduced to
$0.15$ eV. Indeed, the samples are seen in experiments to be dark (Fig. \ref{fig:exp-phases-51}).



In Fig.  \ref{fig:VLPDoS} the calculated DOS for phase V at a
pressure of 36 GPa is shown. The DOS for phase V at 58 GPa is
shown Fig. \ref{fig:VHPDoS}.
A very narrow, singly occupied, d-level is separated from a high binding energy
manifold. A wide insulating gap separates the occupied and unoccupied
d-orbitals. Experimentally the samples seem to be darker, almost black. It
should be noted that the sample contains a mixture of phase IV and V at this
point, and our calculations indicate that the gap in phase IV should be very
small, making phase IV appear black. Phase V has in contrast a larger gap and
should thus be transparent. In the sample the black phase IV is likely obscuring
the transparent phase V.

\begin{figure}
    \includegraphics[width=\linewidth]{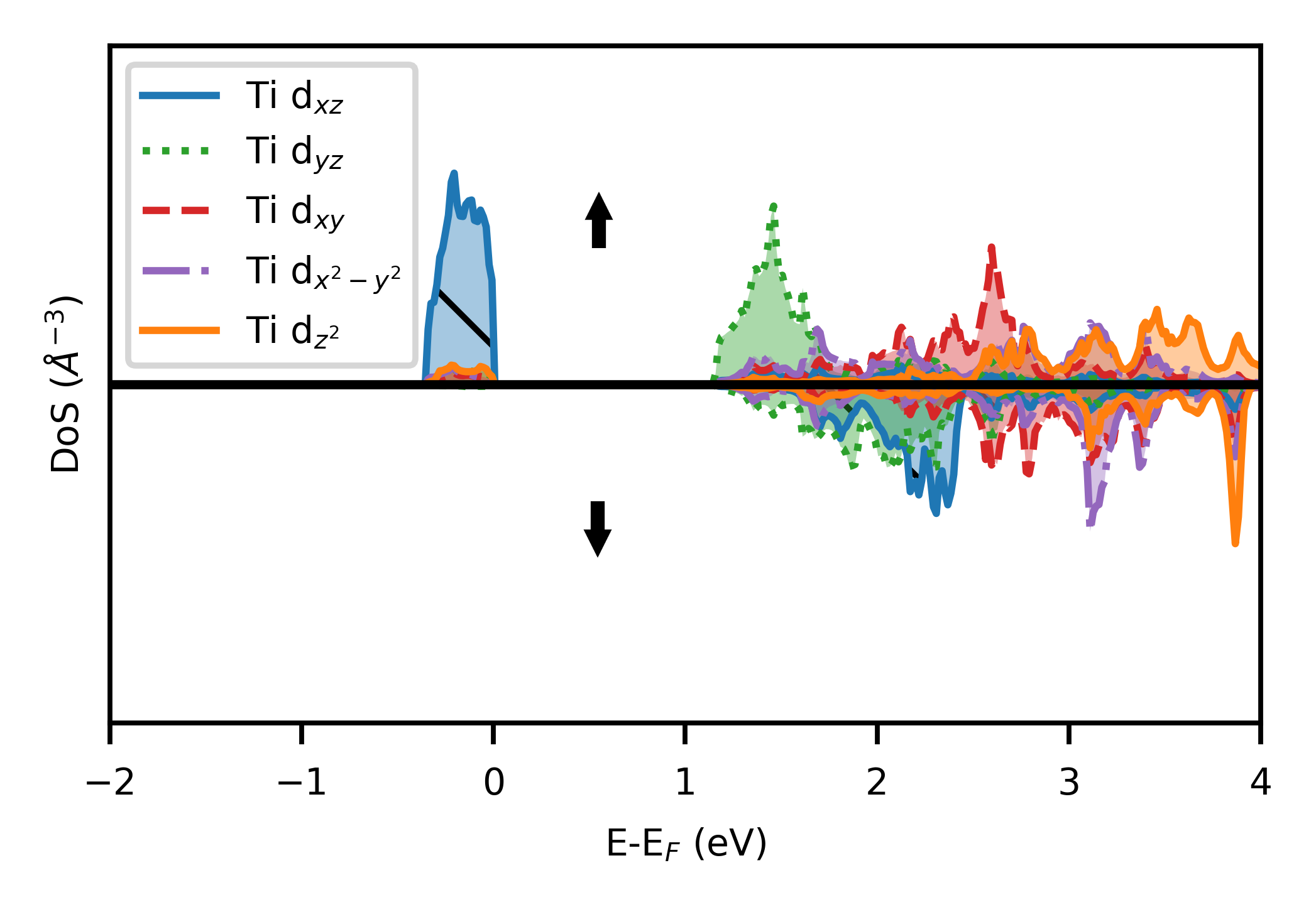}
    \caption{Calculated spin-resolved DOS for one Ti atom in phase V, at 36 GPa.}
    \label{fig:VLPTiDoS}
\end{figure}

The local DOS for one Ti atom in phase
 V is shown in Fig. \ref{fig:VLPTiDoS}. Due to the pentagonal bi-pyramidal oxygen complex
 surrounding the Ti
 atom, the degeneracy of the different d-orbitals is lifted.
 The occupied d-orbitals are seen to be of d$_{xz}$ character, which is separated
  from the the unoccupied d$_{yz}$ orbital, pointing between O atoms. It should be noted that the local
  d$_{xz}$ orbitals of Ti nearest neighbors do not overlap to the same extent
  as in phase IV.

  The gap between occupied and unoccupied d-states appear only if a nonzero on-site
  U is included in the calculation. The separation of d orbitals into upper and
   lower Hubbard bands is thus due to strong correlation between d-electrons.
  The filled O states are well below the low-binding
  energy d-states. Therefore, phase V may be classified as a true Mott
  insulator\cite{jrn:zaanen85}.

  The magnitude of the gap, will
  depend on the choice of the U-parameter. Nevertheless U$= 2.0$ eV seems to
  adequately reproduce the EOS, and simultaneously reproduce the
  insulating state of phase III (and I).
  The gap is intact at the volumes where these phases were observed.
  Thus, the qualitative result of a
  non-metallic state reappearing in the high pressure phases is not sensitive to
  the specific value of U.

  The results strongly suggests that the structural transition is connected with the Mott transition.
  The transformation into a metallic state destabilizes phase III, leading to phases IV
  and V, which do have Mott gaps.

\begin{figure}
    \includegraphics[width=1.3\linewidth,trim={8cm 0 0 0},clip]{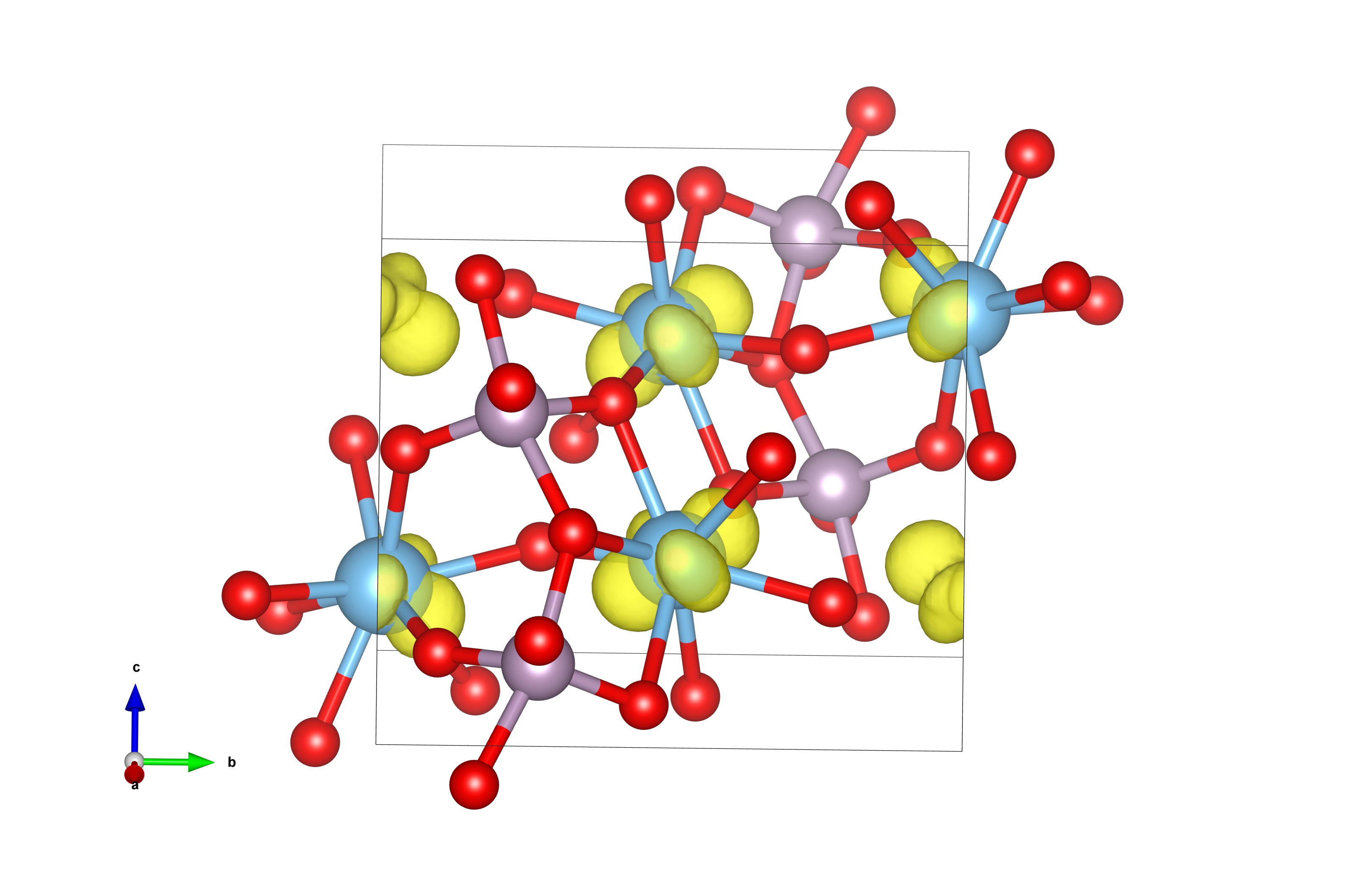}
    \caption{Partial charge density of Ti-d electrons in phase V, at 60 GPa.
    The local $\bar{x}$, $\bar{y}$ and $\bar{z}$-axes have been chosen so that the $\bar{z}$ and $\bar{x}$-axes point
    towards oxygen atoms.}
    \label{fig:VHPParChg}
\end{figure}


Early experimental studies\cite{jrn:kinomura82}  of the magnetic properties of TiPO$_4$
reported a local magnetic moment on Ti$^{3+}$ of 0.8$\mu _B$ at ambient pressure. This is only slightly lower
than what is expected from an insulator with Ti d$^1$ configuration.
By means of \emph{ab-initio} calculations, Lopez et al. \cite{jrn:lopezmoreno12}
reported the magnetic moment to vanish at 12 GPa in the CrVO$_4$ structure
(phase I).
Our calculations indicate that close to the collapse of phase III, the local
magnetic moment is effectively zero.

\begin{figure}
    \includegraphics[width=\linewidth]{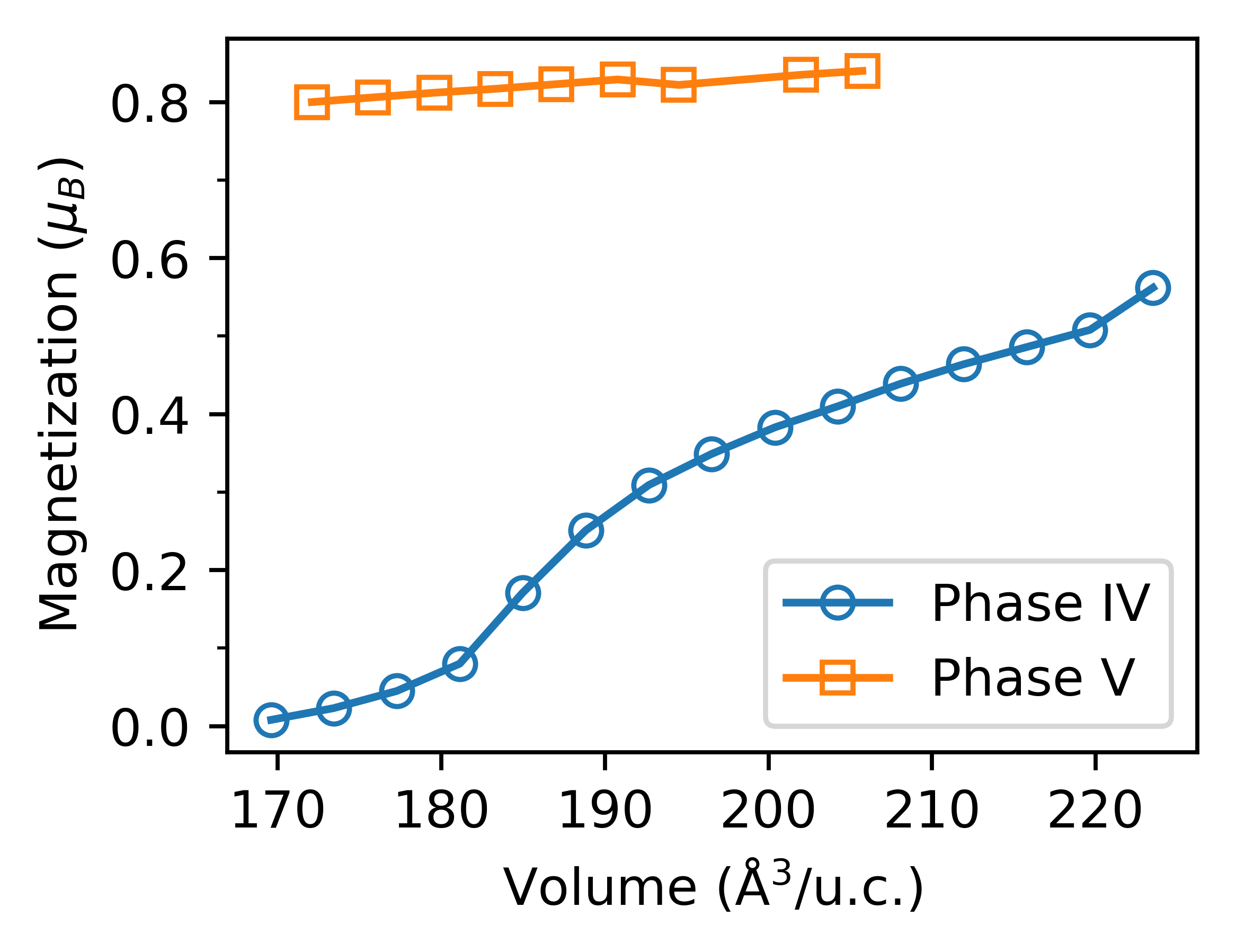}
    \caption{Calculated local magnetic moments as a function of unit cell volume.
    For phases IV and V in AFM configuration.}
    \label{fig:magmom}
\end{figure}

The calculated
magnetic moments for phases IV and V are shown in Fig. \ref{fig:magmom}
as a function of unit cell volume.  At the volume where phase IV becomes stable
 the local magnetic moments are small, $0.3\mu_B$,
and continuously decrease with pressure. At 65 GPa they have practically
vanished, which is also indicated by the depolarized d-level DOS in Fig.
\ref{fig:IVHPTiDoS}.
On the contrary, in phase V the magnetic
moments are comparatively large, $0.8\mu_B$, and decrease very slowly with
increasing pressure.
Due to the crystallographic geometry, there is no overlap between neighboring
d$_{xz}$ states (Fig. \ref{fig:VHPParChg}). The AFM order therefore seems to be due to indirect Ti-O-Ti exchange
coupling.
Phase V is thus clearly antiferromagnetic, and the Ti atoms feature a robust local
magnetic moment, compatible with the d$^1$ configuration.

\section{\label{sec:Disc}Conclusions}\noindent
Calculations show that the recently discovered phase V of TiPO$_4$, featuring PO$_5$
polyhedra, is the energetically favorable structure at high pressure, as
compared to phases III and IV. The less dense phase IV is most likely kinetically stabilized.
These new phases appear when the pressure approaches values at which the known phase III should become
metallic and lose its local magnetic moment. Phase IV shows a larger band gap, but otherwise  behaves similarly to phase III.
In contrast, phase V shows a band gap, and in phase V
magnetism reappears. These results agree well with the color changes observed in
experiments, though the small gap of phase IV obscures the transparent phase V
in the experimental sample containing a mixture of the two phases.

Our finding is in many ways remarkable, as pressure-induced structural transitions in Mott
insulators are usually connected with the disappearance of the band gap.
In phase V of TiPO$_4$, the gap along with magnetism reappear upon a structural
transition. The density of phase V is also higher than that of phases III or IV
and our calculations show that phase V is the stable phase at high pressures.
Thus, it should be possible to detect the inverse pressure-induced Mott
transition in TiPO$_4$ experimentally. In summary,  our study shows that TiPO$_4$
displays intriguing phenomena not only at low temperature but also at high
pressure.

\section*{Acknowledgements}\noindent
This project is funded by the Knut and Alice Wallenberg Foundation for the
project Strong Field Physics and New States of Matter (Grant No. KAW-2013.0020).
We are grateful to the Swedish e-Science Research Centre for financial support.
I.A.A. gratefully acknowledges the Swedish Research Council (VR) grant No. 2015-04391 and the Swedish Government Strategic Research Area in Materials Science on Functional Materials at Link\"oping University (Faculty Grant SFO-Mat-LiU No. 2009 00971). Theoretical analysis of computational results was supported by the Russian Science Foundation (Project No. 18-12-00492).
The computations were performed on resources provided by the Swedish National
 Infrastructure for Computing (SNIC) at High Performance Computing Center North
 (HPC2N) and National Supercomputer Centre (NSC).

\bibliographystyle{apsrev4-1}
\bibliography{p-iv}
\end{document}